\pdfoutput=1

\documentclass[12pt]{article}
\usepackage[
top = 2.5cm, 
bottom = 2.5cm,  
left = 2.55cm,  
right = 2.55cm]{geometry}

\usepackage{amsfonts,amssymb,amsmath,amsthm,cite}
\usepackage{graphicx}
\usepackage{latexsym} 
\usepackage{verbatim}
\usepackage{tikz}
\usepackage{color}
\usepackage{graphicx,amssymb,amsfonts,amsmath,amssymb,amscd,amstext, mathrsfs}
\usepackage{graphicx} 
\usepackage{bbm}
\usepackage{dsfont}
\usepackage{xcolor}
\usepackage[colorlinks=true, citecolor=black, linkcolor=black, allcolors=black]{hyperref}   
\usepackage{amsmath}
\usepackage{empheq}
\usepackage{hyperref}
\usepackage{slashed}
\usepackage{cite}

\newlength\dlf

\newcommand{\MM}{\mathcal{M}}
\newcommand{\ii}{\mathrm{i}}
\newcommand{\np}{{\mathbf{n}^+}}
\newcommand{\nm}{{\mathbf{n}^-}}
\newcommand{\nn}{\mathbf{n}}

\DeclareFontShape{OT1}{cmr}{mx}{n}{<->cmr10}{}
\newcommand{\titlefont}{\fontseries{mx}\selectfont}

\begin{document}

\begin{titlepage}

\begin{flushright} 
\end{flushright}

\begin{center} 
\vspace{1cm}

{\fontsize{22.5pt}{0pt}{\titlefont{{Interface Conformal Anomalies}}}}

\vspace{1.5cm}  

Christopher P. Herzog$^*$,~Kuo-Wei Huang$^+$, and Dmitri V. Vassilevich$^\#$
\\ 

\vspace{1.2cm}
 
{\it
$^*$Department of Mathematics, King’s College London, \\
The Strand, London WC2R 2LS, UK

\vspace{0.5cm}

$^+$Department of Physics, Boston University, \\
Commonwealth Avenue, Boston, MA 02215, USA

\vspace{0.5cm}

$^\#$CMCC, Universidade Federal do ABC, Santo Andr\'e, S.P., Brazil \\ 
and Physics Department, Tomsk State University, Tomsk, Russia
}

\end{center}
\vspace{2cm} 

{\noindent 
We consider two 
$d \geq 2$
conformal field theories (CFTs) glued together along a codimension one conformal interface. The conformal anomaly of such a system  contains both bulk and interface contributions. 
In a curved-space setup, we compute the heat kernel coefficients and interface central charges in free theories. The results are consistent with the known boundary CFT data via the folding trick.  
In $d=4$, two interface invariants generally allowed as anomalies turn out to have vanishing interface  charges. These missing invariants are constructed from components with odd parity with respect to flipping the orientation of the defect. 
We conjecture  that all invariants constructed from components with odd parity may have vanishing coefficient
for symmetric interfaces, even in the case of interacting interface CFT. 
}

\end{titlepage}

\addtolength{\parskip}{0.8ex}
\jot=2 ex

\subsection*{1. Introduction: why interfaces?}

We study boundaries and defects in conformal field theory (CFT) because they have broad applications and because they
are fundamental to our understanding of CFT and quantum field theory more generally.  Boundaries and defects are generally present in most experimental realizations of critical systems, and they bring with them the potential for a wide variety of experimentally verifiable consequences, for example surface critical exponents.  Beyond that, however, there are fundamental questions about the classification and ``space'' of quantum field theories that can be answered through a careful study of defects.  While it is often stated that a conformal field theory is defined -- through operator product expansion -- by its local operator spectrum and set of three-point correlation functions, in fact there are often extended operators, such as Wilson lines, which must be included for a proper definition of the CFT (see e.g.\ \cite{Kapustin:2014gua}).  These extended operators carry with them an additional defect interpretation.

The complete classification of conformal defects or, equivalently, universality classes of critical behavior at the interfaces of CFTs remains a challenging open problem. While earlier studies mostly  focused on $d=2$ critical systems \cite{CARDY1989581, Oshikawa:1996dj, LeClair:1997gz, Bachas:2001vj},\footnote{See, however, earlier discussions about $d=4$ $N=4$ super Yang-Mills theory with an interface \cite{DeWolfe:2001pq, Bak:2003jk, DHoker:2006qeo}. } there has been much recent interest in understanding $d>2$ CFTs with boundaries or defects.\footnote{For examples, see recent works on chiral anomalies and index theorem for the interfaces \cite{Vassilevich:2018aqu,Ivanov:2020fsz}, 
a related work on the $\eta$ invariants \cite{Fukaya:2019qlf}, a study on 't Hooft anomalies with boundaries \cite{Jensen:2017eof}, as well as some physical effects induced by boundary anomalies \cite{Chu:2018ksb}. For a recent review, see \cite{Andrei:2018die}.}

For CFTs without a boundary, the trace anomaly coefficients -- which we call central charges -- provide a useful classification \cite{Deser:1993yx}.  To organize CFTs in the presence of general defects, it is thus natural to look for similar central charges. In this note, we consider a particularly simple example of a defect: a ``symmetric interface''.  A symmetric interface, for us, is a codimension one surface on either side of which we find the same quantum field theory in curved space-time with the same couplings.  Moreover, the dynamical quantum fields are assumed to be continuous across the interface although their normal derivatives may jump.   
The metric is taken continuous across the interface, but is allowed to be otherwise arbitrary and non-smooth.  
Apart from technical interest, such geometries appear in various physical models. An example is the brane-world scenario where the normal derivative of the metric jumps on the brane due to the presence of classical matter. Another example is graphene with a fabricated singular surface; for example, one could glue a cylindrical surface to a disc.\footnote{%
One may consider even more singular geometries, for instance, the electromagnetic fields near the interface of two dielectrics with different permittivities. The effective metric seen by the photons is discontinuous. Due to technical difficulties, we shall not consider such configurations in this work.} 

Our restriction to ``symmetric interfaces'' rules out some important examples.  
Indeed, consider a conformally coupled scalar field that is given a large mass on one side of the interface.  In the limit the mass becomes infinite, one recovers a BCFT with Dirichlet boundary conditions for the scalar.  Even though the scalar is continuous at the interface, the condition to have equal couplings is violated.  
Indeed, more generally it is possible to treat a BCFT as an interface that joins a nontrivial CFT on one side to a trivial one on the other, but such an example is not ``symmetric''.  Our restriction also eliminates Janus theories, where a marginal coupling changes discontinuously at the interface.  

We will largely focus on $d=4$ free CFT and also present results in $d=2, 3$.  
Our main results will be the trace anomalies of $d=4$ free interface CFT (ICFT).  
The computation relies on the heat kernel technique performed in curved spacetime; see \cite{Fursaev:2011zz} for a review. We focus on {\it identical} free CFTs with spin zero, one half, and one on both sides of the interface.
These results generalize the anomaly computation in boundary CFT (BCFT) \cite{Herzog:2015ioa, Fursaev:2015wpa, Solodukhin:2015eca}, which can be recovered via the folding trick.
(Even though a BCFT cannot be thought of directly as a symmetric interface, the folding trick 
involves a nonlocal redefinition of the fields that, for a symmetric choice of metric,
maps the right side of the interface onto the left one.)
It would of course be nice to compute interface and surface charges for non-free theories.  
The only other techniques we are aware of are holography \cite{Estes:2014hka} and  supersymmetric 
localization \cite{Chalabi:2020iie},
both of which apply to restricted classes of theories.

One might expect that the anomaly structure of ICFT can be fixed by that of BCFT through the folding trick. 
However, the moral of the present work is that the general interface story can be richer. 
An interesting observation is that we find two interface invariants that are consistent with all the requirements to be a part of the anomaly but that have zero coefficient  in $d=4$ free ICFT. 
 We do not have a simple argument for this vanishing.  
We  conjecture  this vanishing remains true for general symmetric interfaces, but it is possible that interactions could generate new non-zero charges.

\subsection*{2. Interface Setup}

Let $\Sigma$ be an interface hypersurface where two manifolds or two parts of the same manifold are glued together along their common boundary. The bulk manifold will be denoted by $\MM$, $\dim\MM=d$.   
 We assume that the metric is continuous across $\Sigma$, but not necessarily smooth. Let us mark in an arbitrary way two sides of $\Sigma$ by $+$ and $-$. Let $\np$ and $\nm$ be unit normals to the boundary pointing to $+$ and $-$ sides, respectively. Let $x^\mu$ be local coordinates on $\MM$. The coordinates on $\Sigma$ will be denoted by $x^j$, $j=1,\dots,n-1$. The induced metric on $\Sigma$ will be denoted by $h_{ij}$.
The extrinsic curvatures of $\Sigma$ defined from two sides of $\Sigma$ 
\begin{equation}
K_{ij}^\pm =\Gamma^{\mathbf{n}^\pm}_{ij} \label{Kij}
\end{equation}
do not need to agree. 

There are two particular cases of the interface geometry which are going to play important roles in this work. The first one is a smooth geometry with the metric and all normal derivatives are continuous across the interface. Since $\np=-\nm$, this implies  $K^+_{ij}=-K^-_{ij}$, in particular. We do not put on the interface any fields which interact with the bulk fields. Thus, in the case of a smooth geometry the interface effectively disappears, and so do all surface contributions to the conformal anomaly. The second case corresponds to reflection invariant interfaces obtained by gluing two identical copies of a manifold with boundary. On can think of a spherical cap as an example. In this case, $K^+_{ij}=K^-_{ij}$. As we shall see below, the spectral problem with symmetric interfaces can be reduced to a sum of two boundary value problems -- the folding trick. Again, this is true only if there are no interacting fields at the interface, or if the interaction with such fields have reflection symmetry.

Let $V$ be some vector bundle over $\MM$. Consider an operator $L$ of Laplace type that acts on smooth sections of this bundle. $L$ can be written as
\begin{equation}
L=-(\nabla^2+E) \ ,\label{Lap}
\end{equation}
where $E$ is a smooth endomorphism (a matrix valued function), and $\nabla_\mu=\partial_\mu+\omega_\mu$ is a covariant derivative. We shall also need a bundle curvature
\begin{equation}
\Omega_{\mu\nu}=\partial_\mu\omega_\nu -\partial_\nu\omega_\mu +[\omega_\mu,\omega_\nu] \ .\label{Omega}
\end{equation}
We do not assume any continuity conditions for $E$, $\omega$ and $\Omega$ on $\Sigma$. To have a well defined spectral problem, one has to impose on sections $\phi$ of $V$ some matching conditions on $\Sigma$. A natural choice is to request that $\phi$ is continuous, but the normal derivative jumps,
\begin{eqnarray}
&&\phi^+=\phi^- 
\ , ~~~(\nabla_\np\phi)^+ + (\nabla_\nm\phi)^-=U\phi \label{cond2} \ .
\end{eqnarray}
The superscripts $\pm$ denote  
the limiting values that various quantities take when they approach $\Sigma$ from different sides.\footnote{One can in principle impose more general linear relations between $\phi^\pm$ and its normal derivatives \cite{Bachas:2001vj,Gilkey:2002nv}.} 
We also define a gauge/diffeomorphism vector
\begin{equation}
B_j:=\omega_j^+ -\omega_j^- \label{Bj}
\end{equation}
as the difference between two connections on $\Sigma$.

Our next step is to define conformal matching conditions for various spins. It is instructive to compare them with conformal boundary conditions adopted in, for instance, \cite{Fursaev:2015wpa}. 

\subsubsection*{{\it Scalars:}}

For a conformally coupled scalar field $\varphi$, the operator $L$ reads
\begin{equation}
L=-\Delta + \xi R \ , \qquad \xi=\frac {d-2}{4(d-1)}  \ .\label{Lsc}
\end{equation}
Thus, 
$E=-\xi R \label{Esc}$
while $\omega$, $\Omega$ and $B$ vanish. 
 Under the Weyl rescaling $g^{\mu\nu}\to \bar g^{\mu\nu}=e^{-2\sigma} g^{\mu\nu}$ the operator $L$ and the field $\varphi$ transform as
\begin{equation}
L\to \bar L=e^{-\frac{d+2}2 \sigma}Le^{\frac{d-2}2 \sigma}\ ,\qquad \varphi \to \bar \varphi =e^{-\frac{d-2}2 \sigma}\varphi \  .\label{WLvarphi}
\end{equation}
Also,
\begin{equation}
\bar \nn^\mu=e^{-\sigma}\nn^\mu, ~~~\bar K_{ij}=e^\sigma(K_{ij} - g_{ij}\partial_\nn \sigma) \ , ~~~\bar K=e^{-\sigma} (K-(d-1)\partial_\nn \sigma) \ . \label{WK}
\end{equation}
It is easy to check that the conditions 
 (\ref{cond2}) with 
\begin{equation}
U=\frac{d-2}{2(d-1)} (K^+ +K^-) \label{Usc}
\end{equation}
are Weyl invariant.   
The Euclidean action with an interface is   
\begin{equation}
I=  {1 \over  2 } \int_{\MM / 
\Sigma} d^dx \sqrt{g}  \Big( (\partial \phi)^2 +\frac{d-2}{4(d-1)} R\, \phi^2 \Big) +
{1 \over  2 }  \int_{\Sigma} d^{d-1}x\sqrt{h} U \phi^2  \ .
\end{equation}
The interface contribution is introduced to make the variational problem self-consistent.

\subsubsection*{{\it Spinors:}}

 The massless Dirac operator reads
\begin{equation}
\slashed{D}=\ii \gamma^\mu (\partial_\mu +\omega^{[s]}_\mu) \ , \qquad \omega^{[s]}_\mu = \tfrac 14 w_\mu^{AB}\gamma_A\gamma_B \ .\label{Dirop} 
\end{equation}
Here $A,B,\dots$ are flat indices. Further, $\{ A,B,\dots \}=\{ a,b,\dots, n\}$ so that $e_j^n=0$ and $e_\np^n=1$ on the + side of $\Sigma$ and $e_\nm^n=-1$ on the $-$ side. This implies $\gamma^n=\gamma^\np =-\gamma^\nm$. We have
\begin{equation}
(w_j^{an})^\pm = \mp K_j^{\pm a} \label{wjan} \ .
\end{equation}
In this case, 
\begin{equation}
L=\slashed{D}^2,\qquad   E=-\tfrac 14 R \ ,\qquad \omega=\omega^{[s]},\qquad \Omega_{\mu\nu}=\tfrac 14 \gamma^A \gamma^B R_{AB\mu\nu} \ .\label{EoOsp}
\end{equation}
As $\slashed{D}$ is a first order operator, the matching condition $\psi^+=\psi^-$ implies $(\slashed{D}\psi)^+=(\slashed{D}\psi)^-$. The condition (\ref{cond2}) then yields 
\begin{equation}
U=\tfrac 12 (K^++K^-) \ .\label{Usp}
\end{equation} 
The Weyl invariance of these conditions can be easily checked. We shall need also
\begin{equation}
B_j=\tfrac 12 (K^+_{jb}+K^-_{jb})\gamma^n\gamma^b  \ .\label{Bjsp}
\end{equation}  
The bulk Dirac action is standard and the interface theory does not require a surface term.  

\subsubsection*{\it{$U(1)$ gauge field:}}

The matching conditions for abelian gauge fields and ghosts in the Lorentz gauge are particular cases of matching conditions for the de Rham complex.\footnote{See, for instance, \cite{Gilkey:2001mj} for a more general discussion.}  
For the ghosts ($0$-forms),
\begin{equation}
 L_{\rm gh}=-\Delta \ ,\qquad E=0 \ ,\qquad \omega=0 \ ,\qquad U=0 \ . \label{Lgh}
\end{equation}
Thus, the ghosts are actually smooth across $\Sigma$.  
For 1-forms, the operator $L$ is the Hodge-de Rham laplacian. We have
\begin{equation}
E_A^{\ \ B}=-R_A^{\ \ B}\ ,\qquad \omega_\mu = w_\mu \ , \qquad (\Omega_{\mu\nu})_A^{\ \ B}=-R^B_{\ \ A\mu\nu}\label{EoO1}
\end{equation}
and 
\begin{eqnarray}
&&(B_j)_a^{\ \ n}=-(K^+_{ja}+K^-_{ja})=-(B_j)_n^{\ \ a}\ ,  \\
&& U_a^{\ \ b}=K^+_{ab}+K^-_{ab} \ , ~~ U_n^{\ \ n}=K^++K^- \label{Uvec} \ .
\end{eqnarray} 
Gauge transformed vector fields satisfy matching conditions if the gauge parameter satisfies the matching condition of ghosts.
The gauge invariance of these matching conditions follows by the construction and may be checked directly.   
The bulk gauge-field action is standard and the interface theory does not require a surface term.

\subsection*{3. Heat kernel coefficients and central charges}

For any generalized Laplacian $L$, there is a small-$t$ asymptotic expansion,
\begin{equation}
\mathrm{Tr} \left( f e^{-tL} \right) \simeq \sum_{k=0}^\infty t^{(k-d)/2} a_k(f,L) \ ,  
\label{asym}
\end{equation}
where $f$ is a smearing function which allows a computation of local heat kernel coefficients:
\begin{equation}
a_k(f,L)=\int_\MM d^d x\, \sqrt{g}\, f(x)a_k(x;L) \ . \label{akfL} 
\end{equation}
Note that if there are boundaries or singular surfaces, $a_k(x,L)$ contains $\delta$-functions and derivatives of $\delta$-functions localized on the boundary or on the singular surface.
The trace anomaly is given by
\begin{equation}
\langle T_\mu^\mu (x) \rangle =\eta  \, a_d(x,L) \ , \label{Tan}  
\end{equation}
where $\eta=1$ for bosons and $\eta=-1$ for fermions.

Here we compute the heat kernel coefficients $a_d(f,L)$ ($d=\dim\MM$) by using general expressions. Covariant derivatives are denoted by a semicolon. By a colon we shall denote covariant derivatives on $\Sigma$ containing the Christoffel symbol corresponding to the induced metric $h_{ij}$.\footnote{We follow notation in \cite{Gilkey:2001mj}. Note the Riemann tensor in \cite{Gilkey:2001mj} has an overall minus sign as compared to our notation. The conventions for the Ricci tensor and scalar curvature are identical.}   
 The heat kernel coefficients are local. This means that they are given by integrals of local polynomials. In $a_d(f,L)$, the integral over $\MM$ contains invariants of the canonical dimension $d$, while the integral over $\Sigma$ contains invariants of the dimension $d-1$. 

Before starting actual computations, let us list several consistency conditions \cite{Bordag:1999ed,Gilkey:2001mj} that must be satisfied:

\begin{itemize}
\item[(i)] The heat kernel coefficients have to be invariant with respect to exchanging the roles of the ``+" and ``$-$" sides of $\Sigma$.  The coefficients need to be invariant with respect to which direction one looks at the system.
We expect this invariance to hold for symmetric interfaces more generally.
\item[(ii)] 
When the metric is smooth, we have $K_{ij}^+=-K_{ij}^-$, $R^+=R^-$, etc. 
In this case, the interface effectively disappears and only the bulk structure survives.
\item[(iii)] Assume that $\MM$ is composed of two identical manifolds $\MM^+=\MM^-$ glued together along their common boundary $\Sigma$. Let $x\in \MM^+$ and $x_*\in \MM^-$ be corresponding points. By considering 
\begin{equation}
\phi_{\rm even/odd}(x)=\frac{1}{\sqrt{2}} \left(\phi(x)\pm \phi(x_*) \right) \ ,
\end{equation}
 one can map the heat kernel on $\MM$ with an interface $\Sigma$ to the heat kernel of boundary value problems on $\MM^+$. In our case, this mapping implies that the heat kernel coefficient $a_k$ for the conformal scalar on $\MM$ is a sum of the coefficients $a_k$ for a conformal scalar on $\MM^+$ with Dirichlet boundary conditions on $\Sigma$ and for another scalar with conformal Robin boundary conditions. The heat kernel expansion for a spinor field on $\MM$ has interface coefficients which are twice that for 
conformal spinor fields on $\MM^+$. Similarly, the heat kernel for a $d=4$ Maxwell field on $\MM$ is a sum of the heat kernel for Maxwell fields on $\MM^+$ satisfying the so-called absolute and relative boundary conditions.
\end{itemize}

The computations in two and three dimensions are simple. For the 
conformally coupled scalar we have
\begin{eqnarray}
&&a_2=\frac 1{24\pi} \left[ \int_\MM d^2x\sqrt{g} \, fR +\int_\Sigma dx\sqrt{h}\, 2f(K^++K^-) \right]  \ ,\label{a2sc}\\
&&a_3=\frac 1{1024\pi} \int_\Sigma d^2x \sqrt{h} \bigl( -f(K^++K^-)^2+2f(K^+_{ij}+K^-_{ij})^2 \nonumber\\
&&\qquad\qquad\qquad\qquad +2(K^++K^-)(f_{;\np}+f_{;\nm})\bigr)  \ .\label{a3sc}
\end{eqnarray}
For the Dirac spinor, 
\begin{eqnarray}
&&a_2=-\frac 1{24\pi} \left[ \int_\MM d^2x\sqrt{g} \, fR +\int_\Sigma dx\sqrt{h}\, 2f(K^++K^-) \right] \ , \label{a2sp}\\
&&a_3=\frac 1{512\pi} \int_\Sigma d^2x \sqrt{h} \bigl( f(K^++K^-)^2-2f(K^+_{ij}+K^-_{ij})^2 \nonumber\\
&&\qquad\qquad\qquad\qquad -2(K^++K^-)(f_{;\np}+f_{;\nm})\bigr) \ . \label{a3sp}
\end{eqnarray}

The expressions (\ref{a2sc}) - (\ref{a3sp}) could have been obtained by using only the conditions (i), (ii) and (iii). 
The $d=2$ bulk integral is well known.  
The anomaly in $d=3$ CFT is a surface term.
On $\Sigma$, the expressions $f(K^+-K^-)$ in $a_2$ as well as 
$ f(K^+ - K^-)(K^++K^-)$, $f(K^+_{ij} - K^-_{ij})(K^{+ij}+K^{-ij})$, $(K^+-K^-)(f_{;\np}+f_{;\nm})$ and $(K^++K^-)(f_{;\np}-f_{;\nm})$  
 in $a_3$ are forbidden since they do not satisfy the condition (i).  
The expressions 
\begin{eqnarray} 
f(K^+-K^-)^2\ , ~~~f(K^+_{ij}-K^-_{ij})^2 \ ,  ~~~ (K^+-K^-)(f_{;\np}-f_{;\nm})
\end{eqnarray} 
do not vanish on smooth geometries and thus are forbidden by the condition (ii). The remaining invariants are exactly the ones which appear in (\ref{a2sc})-(\ref{a3sp}); they can be determined by comparing to BCFTs \cite{Solodukhin:2015eca}, as described in (iii) above.

Let us turn to four dimensions.
The bulk contributions are standard. However, here we write them down explicitly with the total derivatives in $\langle T_\mu^\mu  \rangle$ which are sometimes neglected. 
We have
\begin{eqnarray}
&&a_4^{\MM}|_{s=0} = \frac 1{360 (4\pi)^2} \int_\MM d^4x\sqrt{g} \, f \Big( 12(1-5\xi)R_{;\mu}^{\ \ \mu} +5R^2(1-12\xi +36\xi^2) \label{a4sc1} \\
&&\qquad\qquad\qquad \qquad\qquad \qquad  \qquad  \qquad  -2R_{\mu\nu}R^{\mu\nu} +2 R_{\mu\nu\rho\sigma}R^{\mu\nu\rho\sigma} \Big)  \ , \nonumber\\ 
&&a_4^{\MM}|_{ s={1\over 2}} = \frac 1{360 (4\pi)^2} \int_\MM d^4x\sqrt{g} \, f \bigl( -
12R_{;\mu}^{\ \ \mu} +5R^2 -8R_{\mu\nu}R^{\mu\nu} -7 R_{\mu\nu\rho\sigma}R^{\mu\nu\rho\sigma} \bigr) \ , \label{a4sp1} \\
&&a_4^{\MM}|_{s=1}= \frac 1{360 (4\pi)^2} \int_\MM d^4x\sqrt{g} \, f \bigl( -36R_{;\mu}^{\ \ \mu} -50R^2 +176R_{\mu\nu}R^{\mu\nu} -26 R_{\mu\nu\rho\sigma}R^{\mu\nu\rho\sigma} \bigr)  \ .\label{a4em1}
\end{eqnarray}
The parameter  $\xi = \tfrac 16$ corresponds to the conformal scalar, and $\xi=0$  to the ghost. 
We have removed ghost contributions in the electromagnetic field case.

Next, we find the following interface contributions: 
\begin{equation}
a_4^\Sigma = \frac 1{360 (4\pi)^2} \int_\Sigma d^3x\sqrt{h} \,  \sum_{i}  \gamma_i {\cal I}_i
\end{equation}
where  curvature structures ${\cal I}_i$ and heat kernel coefficients $\gamma_i$ are given by
\begin{equation}
\begin{array}{lcccc}
&\ \mbox{ghost}\ &\ \phi \ & \ \psi \ &\  A_\mu\ \\
{\cal I}_1:~f(K^+_{ij}-K^-_{ij})^2(K^++K^-) & -1 & -1 & -4 & -2 \\
{\cal I}_2:~f(K^+_{ij}+K^-_{ij})(K^+_{ij}-K^-_{ij})(K^+-K^-) & -1 & -1 & -4 &  -2 \\
{\cal I}_3:~f(K^+_{ij}-K^-_{ij})(K^+_{jk}-K^-_{jk})(K^+_{ki}+K^-_{ki}) & 2 & 2 & 8 & 4 \\
{\cal I}_4:~f(K^++K^-)^3 & \tfrac {40}{21} & \tfrac {22}{63} & \tfrac{34}{21} & -\tfrac{676}{21}\\
{\cal I}_5:~f(K^+_{ij}+K^-_{ij})^2(K^++K^-) & -\tfrac 47 & -\tfrac {18}7 & \tfrac{26} 7 & \tfrac{580}7\\
{\cal I}_6:~f(K^+_{ij}+K^-_{ij})(K^+_{jk}+K^-_{jk})(K^+_{ki}+K^-_{ki}) & \tfrac {68}{21} & \tfrac {68}{21} & -\tfrac{232}{21} &-\tfrac{872}{21} \\
{\cal I}_7:~(K^++K^-)(K^+-K^-)(f_{;\np}-f_{;\nm}) & -5 & 0 & 10 & 20\\
{\cal I}_8:~(K^+_{ij}-K^-_{ij})(K^+_{ij}+K^-_{ij})(f_{;\np}-f_{;\nm}) & -1 & -1 & -4 & -2\\
{\cal I}_9:~(K^++K^-)^2 (f_{;\np}+f_{;\nm}) & -\tfrac{12}7 & -\frac{29}{21} & \frac{36}7 & \tfrac{60}7 \\
{\cal I}_{10}:~ (K^+_{ij}+K^-_{ij})^2 (f_{;\np}+f_{;\nm}) &\tfrac {18}7 & \tfrac{18}7 & -\tfrac{54}7 & -\tfrac 67 \\
{\cal I}_{11}:~(K^++K^-)(f_{;\np\np}+f_{;\nm\nm}) & 12 & 2 & -12 & -36\\
{\cal I}_{12}:~f(K^++K^-)_{:jj} & 24 & 4 & -24 & -72 \\
{\cal I}_{13}:~f(R^+_{ijkj}-R^-_{ijkj})(K^+_{ik}-K^-_{ik}) & -2 & -2 & -8 & -4\\
{\cal I}_{14}:~f(R^++R^-)(K^++K^-) & 10 & 0 & 10 & -100 \\
{\cal I}_{15}:~f(R^+_{i\np i \np}+R^-_{i\nm i \nm})(K^++K^-) & -2 & -2 & -8 & 176 \\
{\cal I}_{16}:~f(R^+_{i\np j \np}+R^-_{i\nm j \nm})(K^+_{ij}+K^-_{ij}) & 6 & 6 & -36 & 72\\
{\cal I}_{17}:~f(R^+_{ijkj}+R^-_{ijkj})(K^+_{ik}+K^-_{ik}) & -2 & -2 & -8 & 176 \\
{\cal I}_{18}:~f(R^+_{;\np}+R^-_{;\nm}) & 12 & 2 & -12 & -36 \\
{\cal I}_{19}:~(R^+ - R^-)(f_{;\np}-f_{;\nm}) & -5 & 0 & 10 & 20\\
{\cal I}_{20}:~(R^+_{i\np i \np} - R^-_{i\nm i \nm})(f_{;\np}-f_{;\nm}) & -2 & -2 & -8 & -4
\end{array}   
\end{equation}
These surface contributions are computed with the help of the basis considered in \cite[Theorem 7.1]{Gilkey:2001mj} but we remark that \cite{Gilkey:2001mj} uses an overcomplete basis. 
 Using Gauss-Codazzi equations, we can write 
\begin{equation}
{\cal I}_{13}= -{1\over 2} ({\cal I}_{1}+ {\cal I}_{2})+ {\cal I}_{3} \; \; \; \; \mbox{and} \; \; \;\; \;
 {\cal I}_{20} ={1\over 2} ( {\cal I}_7 - {\cal I}_8+{\cal I}_{19})\ .
\end{equation}

\subsubsection*{{\it Interface trace anomaly for $d=4$ ICFTs} }

By collecting everything together, we now obtain the trace anomaly 
\begin{eqnarray}
&&\langle T_\mu^{\ \mu}\rangle =\frac 1{16\pi^2} \left[  ( cW^2 - aE^{\rm bulk})\right. \label{A} \\ && + \left.
\delta(\Sigma) \Big((-a(E^+ + E^-) 
+b_1 \mathrm{tr}(\widehat K^+ + \widehat K^-)^3 
+ b_2 (W_{\nn j \nn k}^+ + W_{\nn j \nn k}^-)(\widehat K^{jk+}+\widehat K^{jk-})  \Big)\right], \nonumber
\end{eqnarray}
where
\begin{eqnarray}
&&E^{\rm bulk}=R^2-4R_{\mu\nu}R^{\mu\nu}+R_{\mu\nu\rho\sigma}R^{\mu\nu\rho\sigma},\nonumber\\
&&E^+=-8R_{j\nn j \nn}^+ K^+ - 8R_{ikjk}^+ K_{ij}^+ +4K^+R^+ + \tfrac 83 K^{+3} +
\tfrac {16}3 K_{ij}^+K_{jk}^+K_{ki}^+ -8K^+K_{ij}^+ K_{ij}^+ ,\nonumber\\
&&\widehat K^+_{ij}=K_{ij}^+-\tfrac 13 h_{ij} K^+ .
\end{eqnarray}
 We have dropped the anomaly $\Box R$ which will be cancelled by the conformal variation of a local counterterm; see (\ref{ct}) below. 
 The central charges are given in the following table: 
\begin{center}
\scalebox{1.1}{
\begin{tabular}{|l|c|c|c|c|}\hline
spin s & $360 a$  &  $360 c$  &  $360b_1$  &  $360 b_2$ \\ \hline
s=0 & $1$  & $3$  &  $\tfrac {32}7$  & 12 \\
s=${1\over 2}$ & 11 & 18  & $\frac {180}7$ &  72  \\
s=1 & 62 & 36 &  $\frac {288}7$  & 144 \\
\hline
\end{tabular}}
\end{center}

Via the standard folding trick, one can check that these central charge results are consistent with BCFT data obtained earlier in \cite{Herzog:2015ioa, Fursaev:2015wpa}. For instance, $\mathrm{tr}(\widehat K^+ +\widehat K^-)^3$ becomes $8~\mathrm{tr}\widehat K^3$ after the folding (see the discussion after (\ref{Kij})); note this is to be compared to $b_1$(Dirchlet)+$b_1$(Robin) in BCFT.  
The relations read
\begin{eqnarray}
b^{\rm{ICFT}}_1= {1\over 4} b^{\rm{BCFT}}_1 \ , ~~~~b^{\rm{ICFT}}_2= {1\over 2} b^{\rm{BCFT}}_2 \ .
\end{eqnarray}
Note $b^{\rm{ICFT}}_2=4c$ in free ICFTs while $b^{\rm{BCFT}}_2=8c$ in free BCFTs. It was shown in \cite{Herzog:2017xha} that such a relation can be violated by boundary marginal interactions in BCFT. 
While we do not include interactions here, we expect the interaction will correct the relation $b^{\rm{ICFT}}_2=4c$.

The integrated trace anomaly is locally conformally invariant.  
We have verified that all the derivative of the delta-function terms can be cancelled by the following local counterterms:
\begin{eqnarray}
&&
I_{ct} =-\frac 1{ (4\pi)^2} \int_\MM d^4x\sqrt{g} ~\alpha_1 R^2
-\frac 1{ (4\pi)^2} \int_\Sigma d^3x\sqrt{h} \left[ \alpha_2 (R^++R^-)(K^++K^-)\right. \nonumber \\
&&~~~~~~~~~~~~~~~~~~~~ \qquad \qquad \left.
+\alpha_3 (K^++K^-)^3 +\alpha_4 (K_{ij}^++K_{ij}^-)^2(K^+ + K^-)\right] \label{ct}
\end{eqnarray}
where  coefficients $\alpha_i$ are given by
\begin{center}
\scalebox{1.1}{
\begin{tabular}{|l|c|c|c|c|}\hline
spin s & $360\alpha_1$ & $360\alpha_2$ & $360\alpha_3$ & $360\alpha_4$ \\ \hline
s=0  & $\tfrac 16 $ & $ \tfrac 13$ & $-\tfrac {44}{189}$ & $\tfrac{6}{7}$ \\
s=${1\over 2}$ & 1 & 2 & $-\frac {10}{21}$ & $\tfrac {18}7$ \\
s=1 & $-3$ & $-6$ & $-\frac {62}{63}$ & $-\tfrac 27$ \\
\hline
\end{tabular}}
\end{center} 
Namely, the conformal transformation of \eqref{ct}, $\delta_\sigma I_{ct}$, reproduces terms in $a_4$ with derivatives of $f$ after replacing $\sigma$ with $f$. 
  
Interestingly,  the requirement of vanishing derivatives of the smearing function $f$ automatically removes the $\Box R$ anomaly in free theories \eqref{a4sc1}, \eqref{a4sp1}, \eqref{a4em1} via the identity $\delta_\sigma \int_\MM R^2= 12 \Box R$. To our knowledge, such a connection to a vanishing $\Box R$ has not been mentioned before. 
Note  central charges $a, c, b_1, b_2$ do not renormalize 
while $\alpha$-terms depend on normalization conditions and thus, in this sense, scheme-dependent.  
However, these $\alpha$-coefficients are  still meaningful as long as one stays in the heat-kernel scheme.

An interface theory with the ${\cal N}=4$ super Yang-Mills multiplet has simple relations:
\begin{eqnarray}
{\rm {\cal N}=4~~SYM:}~~~~~{a\over c}=1\ ,~~~~ {b_2\over b_1}= 3 \ , ~~~~\alpha_1=\alpha_2=0\ ,~~~~{\alpha_4\over \alpha_3}= -3\ .
\end{eqnarray}  The $\Box R$ anomaly has zero coefficient with the ${\cal N}=4$  SYM multiplet.   
More generally, it would be interesting to search for bounds on these coefficients in interface CFT.

\subsection*{4. Concluding Remarks} 

There are two structures,
\begin{eqnarray}
&J_1=\mathrm{tr} \left( (\hat K^+ - \hat K^-)^2 (\hat K^+ + \hat K^-)\right) \delta(\Sigma) \ , \label{J1} \\
&J_2=(W_{\nn j \nn k}^+ - W_{\nn j \nn k}^-) (\widehat K^{jk+}-\widehat K^{jk-})~ \delta(\Sigma)  \ , \label{J2}
\end{eqnarray}
 which satisfy the conditions (i) - (iii) given at the beginning of the Sec. 3 and give rise to conformal invariants after integration. 
However, these invariants never appear in $\langle T_\mu^\mu  \rangle$ for the cases we studied here.  
In particular, they cannot be determined through the folding trick.   

All invariants that appear in (\ref{A}) 
 in the final expression for $\langle T_\mu^\mu \rangle$ can be written as $\mathrm{tr}\, (A\cdot B \dots C)$ with $A$, $B$, $C$ tensors. 
 These tensors are irreducible in the sense that they contain the curvatures in positive powers and cannot be written as products of tensors of a lower dimension.   
We observe that, for allowed invariants, all 
 multiplets
$A$, $B$, etc., are even with respect to the reflection $+\leftrightarrow -$. 
The invariants $J_1$ and $J_2$ instead contain odd factors.  
This observation leads us to conjecture 
the following new rule for symmetric interfaces:
\begin{quote}
{\it 
The allowed interface anomaly must be factorizable 
in irreducible factors which are even under the reflection. 
}
\end{quote}
This rule would allow one to distinguish the invariants which appear in $\langle T_\mu^\mu  \rangle$ from the ones which do not. 
However, we do not know {\it why} it should work generally.  Note  the counterterms (\ref{ct}) follow a similar pattern.

It will be interesting to  
test the conjecture
by including interactions on the interface or in the bulk.  
The boundary trace anomaly  of a graphene-like $d=4$ interacting BCFT was recently discussed in \cite{Herzog:2017xha}. (For supersymmetric generalizations, see \cite{Herzog:2018lqz}.)  
By looking at an interface generalization of this theory, one could test the conjecture.
While this graphene-like theory has interactions confined to the boundary, one could also study what happens with interactions in the bulk, for example by looking at maximally supersymetric $SU(N)$ Yang-Mills theory in the presence of an interface.

It would further be interesting to understand the implication of these results for displacement operator correlators in $d\geq 2$ ICFTs.  
Recall that, in BCFT, the Ward identities hold away from the boundary but there are corrections on the boundary:
\begin{eqnarray}
\partial_\mu T^{\mu {\rm n}}= D(x^{\perp}) \delta(x^{\rm n}) \ ,
\end{eqnarray}
where the displacement operator, ${\cal D}$, is dual to the position of the boundary and plays a universal role in BCFT. 
Similarly, one may define the displacement operator in ICFT as the difference between the normal-component of the stress tensors:
\begin{eqnarray}
{\cal D} \sim  (T^{nn}_+ - T^{nn}_-)|_{\Sigma} \ .
\end{eqnarray}
 In the boundary case, it is known that the coefficients 
of the two- and three-point functions of the displacement operator are proportional to the $b_1$ and $b_2$ central charges \cite{Herzog:2017xha,Herzog:2017kkj}.  
In the interface case, how are the boundary invariants related to the displacement operator?  It seems natural to expect that the relation between $b_1$, $b_2$ and the displacement operator continues to hold.  But then it is not clear to what one should relate the coefficients of the additional invariants (\ref{J1}) and (\ref{J2}).  Perhaps their absence correlates with the absence of corresponding operators on the interface.

It will be also nice to consider the trace anomalies and boundary/interface central charges in $d=5,6$ ICFTs to check if the above conjectured rule applies. 

\subsection*{Acknowledgments}

{\noindent 
We thank K. Jensen and A. Karch for interesting conversations. We also thank S. Solodukhin for correspondence. 
C.H. was supported in part by the U.K. Science \& Technology Facilities Council Grant ST/P000258/1 and by a Wolfson
Fellowship from the Royal Society.
 K-W.H was supported in part by the Simons Collaboration grant on the Non-Perturbative
Bootstrap and in part by the U.S. Department of Energy Office of Science under award
number DE-SC0015845. 
 D.V.V. was supported in parts by the
Sao Paulo Research Foundation, project 2016/03319-6, by the grant 305594/2019-2 of
CNPq, by the RFBR project 18-02-00149-a and by the Tomsk State University Competitiveness
Improvement Program.}

\bibliographystyle{utphys}
\bibliography{icftRefs}

\end{document}